\documentclass[useAMS,usenatbib]{mnras}

\usepackage{graphicx}
\usepackage{grffile}\usepackage{epsf}
\usepackage{amsmath}
\usepackage{amssymb}

\newcommand{\kmsmpc}{\kms\;{\rm Mpc}^{-1}}

\newcommand{\kms}{{\rm km}\,{\rm s}^{-1}}
\newcommand{\cms}{{\rm cm}^{-2}}
\newcommand{\cmc}{{\rm cm}^{-3}}

\newcommand{\msolar}{{\rm M}_{\odot}}

\newcommand{\gad}{{\sc Gadget-3}}
\newcommand{\CIV}{\hbox{C\,{\sc iv}}}

\newcommand{\OVI}{\hbox{O\,{\sc vi}}}

\newcommand{\HI}{{\hbox{H\,{\sc i}}}}

\newcommand{\nh}{{n_{\rm H}}}

\hypersetup{draft} 

\begin{document}
\title[Black holes transform galaxies by ejecting halo gas]{Feedback from supermassive black holes transforms centrals into passive galaxies by ejecting circumgalactic gas} 

\author[B. D. Oppenheimer et al.]{
\parbox[t]{\textwidth}{\vspace{-1cm}
Benjamin D. Oppenheimer$^{1}$\thanks{benjamin.oppenheimer@colorado.edu}, Jonathan J. Davies$^{2}$, Robert A. Crain$^{2}$, Nastasha A. Wijers$^{3}$, Joop~Schaye$^{3}$, Jessica K. Werk$^{4}$, Joseph N. Burchett$^{5}$, James W. Trayford$^{3}$, Ryan Horton$^{1}$}\\\\
$^1$CASA, Department of Astrophysical and Planetary Sciences, University of Colorado, 389 UCB, Boulder, CO 80309, USA\\ 
$^2$Astrophysics Research Institute, Liverpool John Moores University, 146 Brownlow Hill, Liverpool, L3 5RF, UK\\
$^3$Leiden Observatory, Leiden University, P.O. Box 9513, 2300 RA, Leiden, The Netherlands\\
$^4$University of Washington, Department of Astronomy, Seattle, WA, USA\\
$^5$University of California - Santa Cruz, 1156 High Street, Santa Cruz, CA 95064
}
\maketitle

\pubyear{2019}

\maketitle

\label{firstpage}

\begin{abstract}

\citet{davi19a} established that for $L^*$ galaxies the fraction of baryons in the circumgalactic medium (CGM) is inversely correlated with the mass of their central supermassive black holes (BHs) in the EAGLE hydrodynamic simulation.  The interpretation is that, over time, a more massive BH has provided more energy to transport baryons beyond the virial radius, which additionally reduces gas accretion and star formation.  We continue this research by focusing on the relationship between the 1) BH masses, 2) physical and observational properties of the CGM, and 3) galaxy colours for Milky Way-mass systems.  The ratio of the cumulative BH feedback energy over the gaseous halo binding energy is a strong predictor of the CGM gas content, with BHs injecting $\ga 10\times$ the binding energy resulting in gas-poor haloes.  Observable tracers of the CGM, including $\CIV$, $\OVI$, and $\HI$ absorption line measurements, are found to be effective tracers of the total $z\sim 0$ CGM halo mass.  We use high-cadence simulation outputs to demonstrate that BH feedback pushes baryons beyond the virial radius within $100$ Myr timescales, but that CGM metal tracers take longer ($0.5-2.5$ Gyr) to respond.  Secular evolution of galaxies results in blue, star-forming or red, passive populations depending on the cumulative feedback from BHs.  The reddest quartile of galaxies with $M_*=10^{10.2-10.7}\ \msolar$ (median $u-r = 2.28$) has a CGM mass that is $2.5 \times$ lower than the bluest quartile ($u-r=1.59$).  We propose strategies for observing the predicted lower CGM column densities and covering fractions around galaxies hosting more massive BHs using the Cosmic Origins Spectrograph on {\it Hubble}.

\end{abstract}

\begin{keywords}
methods: numerical; galaxies: formation, intergalactic medium, supermassive black holes; cosmology: theory; quasars: absorption lines 
\end{keywords}

\section{Introduction}  
Galaxies residing in Milky Way (MW)-mass haloes display great diversity.  While the typical halo mass of $1-2\times 10^{12}\ \msolar$ hosts a
central galaxy with a stellar mass of several times $10^{10}\ \msolar$
\citep[e.g.][]{beh13a, moster13}, the rate of present-day star
formation (SF) varies by orders of magnitude
\citep[e.g.][]{som08,mous13,henr15}.  This is often discussed in terms
of the ``blue'' SF cloud and the ``red'' passive sequence, and appears
to indicate a process of galaxy transformation sometimes referred to
as ``quenching.''  Revealing the process by which a galaxy's star formation rate
(SFR) is curtailed over a relatively narrow range of halo mass is a key
motivation for exploring sophisticated models of galaxy formation and
evolution, especially cosmologically-based hydrodynamic simulations
that self-consistently follow the gas that fuels SF.

Cosmological hydrodynamic simulations that are now able to reproduce
fundamental properties of galaxy populations and the morphological sequence of
the Hubble Tuning Fork include EAGLE \citep[Evolution and Assembly of
  GaLaxies and their Environments][]{sch15}, Illustris-TNG
\citep{pil18}, Horizon-AGN \citep{dubois16}, and MUFASA \citep{dav16}, the first of which we use
here.  Galaxy formation theory has long predicted a transition at the
$\sim 10^{12}\ \msolar$ dark matter (DM) halo mass above which virial
equilibrium achieves temperatures with lowered baryonic cooling
efficiencies \citep{ree77,sil77,whi78}.  Nevertheless, the cooling
rate in gaseous haloes, now termed the circumgalactic medium (CGM), is
over-efficient in forming stars compared to the stellar fraction
observed in galaxies \citep{whi91, bal01, ker05}.  Simulations apply
feedback mechanisms to reduce the efficiency of accretion onto and
star formation within a galaxy to solve this over-cooling problem.
Earlier hydrodynamic models applied forms of stellar superwind
feedback to eject baryons from the galactic sites of star formation to
overcome this over-cooling problem \citep{spr03b, opp06, sch10}.
While these simulations often reproduced lower mass galaxy properties,
their lack of feedback from super-massive black hole (BH) growth was
considered a possible missing ingredient for preventing continued star
formation and stellar assembly in galaxies at and above MW masses
\citep[e.g.][]{opp10}.

Since its earliest inclusions in cosmological models, BH feedback is
required to become effective at the MW-mass scale to reproduce galaxy
properties \citep[e.g.][]{bow06,cro06,sijacki07,boo09}.  Hence, the
halo mass at which cooling efficiencies decline is similar to the
masses where significant BH feedback turns on.  This may not be a
coincidence as argued by \citet{bow17}, who link the rapid phase of BH
growth to the formation of a hot halo that prevents efficient ejective
stellar feedback.  Using the same EAGLE simulation we explore here,
\citet{bow17} showed that buoyant thermal stellar feedback becomes
inefficient upon the formation of a hot halo at halo mass
$\sim 10^{12}\ \msolar$, concentrating gas in the centre where it
triggers non-linear BH growth and feedback.  The BH feedback imparted
to the surrounding halo effectively curtails star formation and can
transition the galaxy from the blue cloud to the red
sequence. 

As already discussed, the central galaxies hosted by MW-mass haloes
have diverse SFRs, colours, morphologies, and BH masses, $M_{\rm BH}$.
\citet[][hereafter D19]{davi19a} used EAGLE to show that the diversity
extends to their gaseous haloes with the primary astrophysical driver
being $M_{\rm BH}$, which can be thought of as proportional to the
integral of BH feedback energy imparted over its growth history.  The
rapid BH growth phase during which $M_{\rm BH}$ multiplies by factors
of several over a small fraction of the Hubble timescale, also clears
out a significant fraction of the baryons in the CGM.  D19
demonstrated that the scatter in gaseous baryon fractions,
$f_{\rm CGM}$, at a given halo mass is highly anti-correlated with
$M_{\rm BH}$, indicating a causal relation between $f_{\rm CGM}$ and
$M_{\rm BH}$.  This conclusion was confirmed by comparing to a
simulation without AGN feedback.  They also showed that the greatest
scatter of $f_{\rm CGM}$ in EAGLE occurs at halo masses
$M_{200} = 10^{12.0-12.3}\ \msolar$, where $M_{200}$ is the mass within
a sphere within which the mean internal density is $200\times$ the
critical overdensity.  Following D19, we define

\begin{equation} \label{equ:fCGM}
f_{\rm CGM} \equiv M_{\rm gas}(R<R_{200})/M_{200}(R<R_{200}) \times \Omega_{\rm M}/\Omega_{\rm b}
\end{equation}
\noindent where $R_{200}$ indicates the radius of a spherical
$M_{200}$ mass distribution.  Here, $M_{\rm gas}$ includes all CGM gas
regardless of temperature, but excludes ISM gas, which is in contrast
with D19 and leads to only a small decrease.  D19 also showed that
$f_{\rm CGM}$ is highly correlated with SFR, which they argue is a
result of the BH removing CGM gas that provides fuel for further SF.


$f_{\rm CGM}$ is not directly obtainable for galactic haloes with
current observational facilities.  While D19 determined that soft
X-ray emission and thermal Sunyaev-Zel’dovich (S-Z) measurements serve
as effective indicators of the baryon fraction within haloes,
X-ray \citep[e.g.][]{bog13,li17} and S-Z \citep{planck13,greco15}
measurements for a statistically significant sample of MW-mass
galaxies are beyond the capabilities of current instrumentation.  In
contrast, the UV probes of the CGM afforded by the Cosmic Origins
Spectrograph (COS) on the {\it Hubble Space Telescope} provide a
growing database of gaseous halo measurements that can be correlated
with galaxy properties \citep[e.g.][]{stocke13,tum13, burc15,
  tum17}. Studies that combine gas column density measurements from UV
absorption-line spectra with ionization modelling have estimated
$f_{\rm CGM}$ for galaxies with M$_{200}$ $\approx$
10$^{12}$M$_{\odot}$ to range from 25 - 100\%, albeit subject to large
systematic uncertainties \citep[e.g.][]{werk14, pro17}. Our goal is to
identify a more sensitive, empirical indicator of the baryonic content
of gaseous haloes around nearby galaxies, for which we can also obtain
an estimate of $M_{\rm BH}$.


D19 demonstrated that it is the integral of BH feedback, not the
instantaneous BH feedback, that ultimately ejects baryons beyond
$R_{200}$. Specifically, while $f_{\rm CGM}$ strongly anti-correlates
with $M_{\rm BH}$, it does not correlate at a significant level with
the BH accretion rate at $z=0$.  This is supported by the
observational result that COS sightlines intersecting haloes hosting low-redshift AGN show little if any effect on CGM ion column densities \citep{berg18}.

Another of our goals is to explore the relations and timescales
between BH feedback and the clearing of the gas from the CGM.  We
leverage high-cadence EAGLE ``snipshot'' outputs that track galaxies,
BHs, and CGM absorption signatures on $\la 100$ Myr timescales.
\citet{corr19} used the same snipshots to determine a causal link
between AGN activity and the transition of a central galaxy's colour
from the blue cloud into the ``green valley" of intermediate galaxy
colours, and to the red sequence \citep[see also][]{tra16}.  We extend
this type of investigation to the CGM to determine how rapidly baryons
within $R_{200}$ respond, as well as UV absorption line indicators of
$f_{\rm CGM}$.

The final goal of this study is to connect the baryonic content of the
CGM to the processes by which a galaxy becomes a part of the red
sequence or remains in the blue cloud.  From the strong correlation
between $f_{\rm CGM}$ and galactic SFR (D19), we expect a galaxy's
colour to also change in response to the BH.  Therefore we use the
{\tt SKIRT} radiative transfer scheme \citep{camps15} that applies
dust-reddening to EAGLE galaxies \citep{tra17} to consider the
following questions: Can rapid BH growth not only lift the CGM, but
change the colour of a galaxy by curtailing the fuel supply for SF?
Is the CGM of a red galaxy substantially more diffuse and less massive
than that of a blue galaxy in the MW halo mass range?  What is the
relationship between a galaxy's colour, its CGM content probed in the
UV, and its central BH?

The paper is laid out as follows.  We introduce the EAGLE simulation,
define our galaxy samples, and describe the methods to calculate
physical and observational quantities in \S\ref{sec:methods}.  In
\S\ref{sec:CGM_BH}, we revisit the work of D19 linking the CGM gas
fraction to $M_{\rm BH}$ and introduce the ratio of BH
feedback energy relative to the binding energy of the CGM.  We then
discuss available UV indicators for $f_{\rm CGM}$, highlighting $\CIV$
covering fractions, around galaxies in the local Universe in
\S\ref{sec:CIV}.  We access high-cadence outputs to track the CGM in
response to the central BH in \S\ref{sec:cadence} and integrate galaxy
colours into the analysis in \S\ref{sec:colors}.  We discuss how
the $\OVI$ ion traces the CGM baryon content in \S\ref{sec:OVI}, how the
efficiency of BH feedback determines galaxy properties in
\S\ref{sec:BHeff}, and consider our Galaxy's BH and CGM in
\S\ref{sec:MilkyWay}.  We summarize in \S\ref{sec:summary}.

\section{Simulations} \label{sec:methods}

\subsection{The EAGLE simulation and black hole energetics} \label{sec:sims}

We use the main EAGLE ``Reference'' simulation volume that is 100
comoving Mpc on a side, referred to as Ref-L100N1504.  This $1504^3$
DM and smooth particle hydrodynamic (SPH) particle run uses a heavily
modified version of the N-body/Hydrodynamical code \gad~\citep[last described
in][]{spr05}.  The DM particle masses are $9.7\times 10^6\ \msolar$ and
the initial SPH particle masses are $1.8\times 10^6\ \msolar$.  The
simulation was originally published by \citet{sch15} and \citet{cra15}
\citep[see][for the public release]{mcalpine16,cam18}.  EAGLE applies
the pressure-entropy SPH formulation of \citet{hopk13} with a series
of additional SPH implementations referred to as {\sc ANARCHY}
\citep[see Appendix A of][]{sch15}.  EAGLE assumes a \citet{planck13}
cosmology ($\Omega_{\rm m}=0.307$, $\Omega_{\Lambda}=0.693$,
$\Omega_{\rm b}=0.04825$, $H_0= 67.77$ $\kmsmpc$).

The EAGLE code applies a number of subgrid physics modules, which
include radiative cooling \citep{wie09a}, star formation
\citep{sch08}, stellar evolution and metal enrichment \citep{wie09b},
BH formation and accretion \citep{boo09,sch15,ros16}, stellar feedback
\citep{dal12}, and BH feedback \citep{boo09}.  The stellar and BH
feedback schemes both use thermal prescriptions where the imparted
feedback energy heats local SPH particles.  The calibration of these
schemes is described in \citet{cra15}.

The BH (AGN) feedback scheme is critically important for our
calculations of BH feedback energy.  As in \citet{spr05b}, EAGLE
follows BHs from seed particles that have an initial mass of
$10^5 h^{-1}\ \msolar$, where $h=0.6777$.  BH seeds are placed at the
centre of every halo that exceeds $10^{10} h^{-1}\ \msolar$.  As
described in \citet{sch15}, BH particles grow both by mergers with
other BH particles and through gas accretion.  The growth rate of a BH
due to gas accretion is
$\dot{m}_{BH} = (1-\epsilon_{r}) \dot{m}_{acc}$, where $\dot{m}_{acc}$
is the gas accretion rate based on the \citet{bon44} rate and
$\epsilon_{r}$, the radiative efficiency of the accretion disk, is
assumed to be $10\%$.  The energy feedback rate imparted to the
surrounding gas particles is $\epsilon_f \epsilon_r \dot{m}_{acc} c^2$
where the thermal feedback efficiency $\epsilon_f=15\%$ (the other
85\% is assumed to be radiated away with no coupling to the
surrounding gas).  Hence, we calculate a BH energy feedback efficiency based on the BH mass using

\begin{equation} \label{equ:EBH}
E_{\rm BH} = \frac{\epsilon_f \epsilon_r}{1-\epsilon_r} M_{\rm BH} c^2, 
\end{equation}

\noindent which translates to 1.67\% of the rest mass of the BH with
$c$ being the speed of light.  Although seed particles do not
contribute to the feedback energy in the above equation, their
contributions to the mass assembly are small \citep[e.g.][]{boo09},
since the final mass of BHs in MW mass galaxies by $z=0$ is typically
$\gg 10^5 h^{-1}\ \msolar$.

\subsection{Central galaxy samples}

We focus on a subset of MW-mass central galaxies.  Our main sample is
referred to as the ``$M^*$'' sample, which has
$M_*=10^{10.2-10.7}\ \msolar$ in haloes with
$M_{200}=10^{12.0-12.3}\ \msolar$ and contains 514 galaxies in the
$10^6$ comoving Mpc$^3$ volume.  These objects show great diversity in
$f_{\rm CGM}$, $M_{\rm BH}$, and SFR (D19, their fig. 2).  We also
consider all central galaxies that have $M_*=10^{10.2-10.7}\ \msolar$
regardless of $M_{200}$ in the ``$L^*$'' sample, containing 1106
galaxies, because an observational determination of $M_{200}$ is often
unavailable.  We will show that this sample, which additionally
contains 342 (250) galaxies with $M_{200}<10^{12.0}$ ($>10^{12.3}$)
$\msolar$, often produces similar trends as the $M^*$ sample.

Finally, we identify a sample of secularly evolving $M^*$ galaxies
have not experienced a major merger since at least $z=1.487$ according
to the EAGLE merger trees calculated from the $13$ snapshots between
$z=1.487$ and $0.0$.  A major merger is defined as at least a 3:1
ratio merger with another galaxy of at least $M_* = 10^{9.5}\ \msolar$.
This ``secular'' sample contains 246 galaxies.  While this sample
represents a biased sample of $M^*$ galaxies, these galaxies are used
for high-cadence tracking with up to $155$ ``snipshot'' outputs going
back to $z=5$ to capture the influence of the black hole on the galaxy
and its CGM.  This work is similar to \citet{corr19} who used snipshot
outputs to determine the timescale of galaxy transformation to the red
sequence, but our investigation focuses on our refined subset of
galaxies and considers their CGM properties.

\subsection{Descriptions of physical and observational quantities}

\subsubsection{Black hole masses}

We report BH masses as that of the most massive BH within 50 kpc of
the galaxy centre.  Other black holes exist in the haloes, but their
mass is usually far smaller than that of the central black hole in
these haloes where one galaxy dominates the stellar mass.  We also
track $M_{\rm BH}$ for our secular sample at high time cadence in
\S\ref{sec:cadence}, where we correlate changes in $M_{\rm BH}$ with
CGM properties.  While we do not filter this sample for major BH
mergers, we expect the BH growth for the secular sample to be
dominated by gas accretion, given that we filter out galaxy (and their
associated BH) major mergers by definition of this sample.

\begin{figure*}
\includegraphics[width=0.49\textwidth]{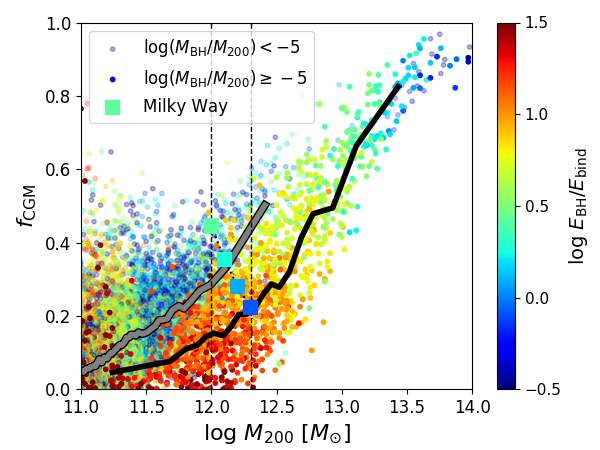}
\includegraphics[width=0.49\textwidth]{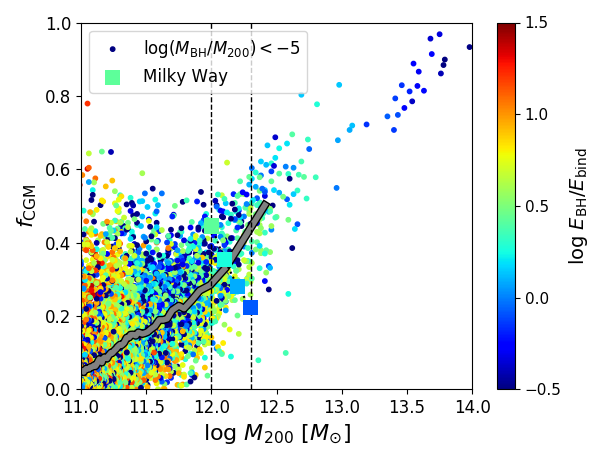}
\caption[]{Halo gas fraction as a function of halo mass in EAGLE,
  coloured by the integrated energy of BH feedback divided by the
  gaseous halo binding energy.  Galaxies with black hole masses $\geq 10^{-5}
  M_{200}$ are plotted in solid (their median is shown by the black
  line), while those with smaller mass are faded and shown alone in
  the right panel (median shown as the grey line).  $E_{\rm BH}/E_{\rm
    bind}$ shows a vertical gradient with $f_{\rm CGM}(M_{200})$ in the galaxy
  halo regime, indicating the integrated feedback from black holes is
  anti-correlated with $f_{\rm CGM}$.  The best estimate for the Milky Way-like halo masses is shown
  shown using large squares for four estimates of halo masses spanning
  our $M^*$ regime from $10^{12.0-12.3}\ \msolar$ bounded by grey dashed lines. } 
\label{fig:m200_fgas}
\end{figure*}

\subsubsection{Absorber column densities}

Ion column densities are calculated by projecting SPH particles onto a
two-dimensional map as described by, e.g., \citet{opp18c}.  
Ionization
fractions for a range of species, including $\HI$, $\CIV$, and $\OVI$
are calculated using CLOUDY-calculated lookup tables as a function of
density, temperature, and redshift.  Calculations assume ionization
equilibrium, and include collisional ionization and photo-ionization
from the time-evolving \citet{haa01} background that was also used for
the calibration of the cooling rates during the simulation.  We use
the self-shielding prescription of \citet{rah13} for $\HI$ column
densities, although this does not affect our covering fraction
calculations.  We include gas only within a radius of
$3\times R_{200}$ from the centre of the galaxy, which separates out
contamination from the CGM of neighbouring galaxies
\citep[e.g.][]{opp18c}.  Three perpendicular projections are used to
calculate covering
fractions.  

\subsubsection{Galaxy colours}

We use the {\tt SKIRT} radiative transfer dust-attenuated colours,
which are available in the data release of \citet{cam18}.  We use the
Sloan Digital Sky Survey (SDSS) $u$ and $r$-band absolute magnitudes
to calculate $u-r$ colours for a random orientation of each galaxy.
We report $u-r$ colours back to $z=2.5$, using their rest-frame
colours.

\section{The dependence of the CGM on the central black hole}  \label{sec:CGM_BH}

We begin this section by casting the large-scale physical parameters
of galactic haloes in terms of integrated feedback energies compared
to gaseous halo binding energies, and then focus on the variety of CGM
and galactic properties of MW-mass haloes.  We also discuss energy
budgets of stellar and BH feedback in terms of CGM gas fractions.

Figure \ref{fig:m200_fgas} plots $f_{\rm CGM}$ as a function of
$M_{200}$.  This was the central plot in D19 (their Figs. 1 and 2),
where they coloured the data points by the scatter in other
parameters, finding a strong anti-correlation between $f_{\rm CGM}$
and $M_{\rm BH}$ at fixed $M_{200}$.  In the left panel, we colour our
data points by the ratio of the integrated feedback energy from the
central BH, $E_{\rm BH}$ from Equation \ref{equ:EBH}, to an estimate
of the binding energy of the gaseous halo $E_{\rm bind}$, where

\begin{equation}
  E_{\rm bind}= \frac{3 G M_{200}^2}{5 R_{200}}\frac{\Omega_{\rm b}}{\Omega_{\rm M}}
\end{equation}

\noindent and $G$ is the gravitational constant.  Note that our
$E_{\rm bind}$ is not the ``intrinsic'' binding energy of the halo
used by D19, which was directly calculated from a paired DM-only
simulation, but an approximation of the energy necessary to unbind a
gaseous halo, hence the $\Omega_{\rm b}/\Omega_{\rm M}$ term.  We use
opaque (faded) circles to indicate whether $M_{\rm BH}/M_{200}$ is
greater (less) than $10^{-5}$.  The black
line shows the running median $f_{\rm CGM}(M_{200})$ value for the
opaque points, while the grey line indicates the median for faded
points having higher gas fractions.  The nearly vertical colour gradient 
of opaque points indicates $f_{\rm CGM}$ depends on $E_{\rm BH}/E_{\rm bind}$
at $M_{200}=10^{12}-10^{13}\ \msolar$ when $M_{\rm BH}/M_{200}>10^{-5}$.  We bracket the
$10^{12.0}-10^{12.3}\ \msolar$ haloes, because this slice shows the
greatest diversity in $f_{\rm CGM}$, which we will later argue is
related to the diverse set of galactic properties of their centrals.


\begin{figure*}
\includegraphics[width=0.49\textwidth]{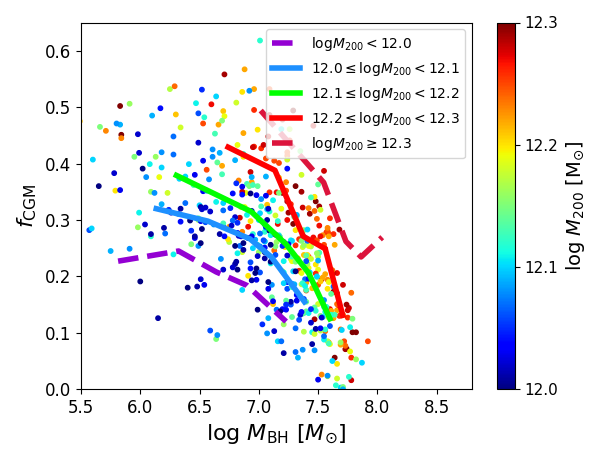}
\includegraphics[width=0.49\textwidth]{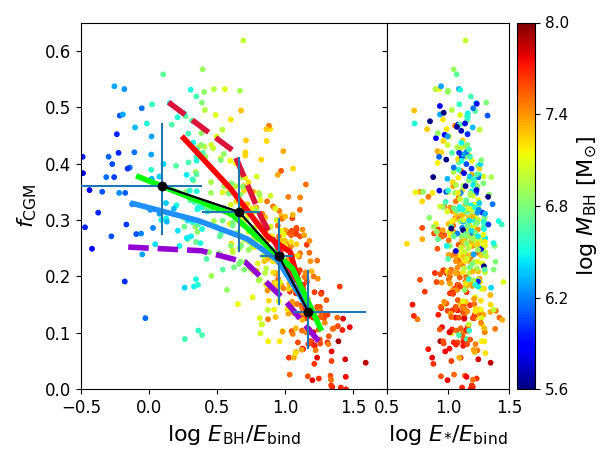}
\caption[]{{\it Left panel:} Halo gas fraction as a
  function of black hole mass for central galaxies with
  $M_*=10^{10.2-10.7}\ \msolar$ and $M_{200}=10^{12.0-12.3}\ \msolar$,
  coloured by halo mass.  We also show median relations in solid lines
  for three 0.1-dex halo bins, and include two dashed lines for all haloes
  hosting $M_*=10^{10.2-10.7}\ \msolar$ galaxies below $10^{12.0}$ and
  above $10^{12.3}\ \msolar$, for which points are not included.  {\it
    Right panel:} The same haloes are plotted as a function of the ratio of the integrated $E_{\rm
    BH}$ over $E_{\rm bind}$, coloured by $M_{\rm BH}$ in the left subpanel.
  The median curves for the different halo mass bins now converge at high $E_{\rm
    BH}/E_{\rm bind}$ values, indicating this relation is more
  fundamental for $f_{\rm CGM}$.  Four quartiles of $E_{\rm BH}/E_{\rm
    bind}$ are shown as black points with error bars indicating
  $1-\sigma$ dispersions for $f_{\rm CGM}$ and the ranges of each
  quartile.  The right subpanel shows the integrated stellar feedback
  energy, which is almost always greater than $E_{\rm BH}$, but does
  not show a correlation with $f_{\rm CGM}$.  }
\label{fig:mBH_fbar}
\end{figure*}

In the left panel, the $M_{\rm BH}/M_{200} < 10^{-5}$ data points 
are difficult to see because they are underplotted, which is why we plot these haloes alone in the right
panel.  These haloes do not show a clear dependence of $f_{\rm CGM}(M_{200})$ on
$E_{\rm BH}$, suggesting that BH feedback becomes efficient at
ejecting CGM only above $M_{\rm BH}/M_{200} \ga 10^{-5}$ 
for MW-mass haloes, i.e. $M_{\rm BH}\ga 10^{7.0-7.3}\ \msolar$, as shown by \citet{bow17}.  

We now focus on our $M^*$ sample, showing data points of $f_{\rm CGM}$
against $M_{\rm BH}$ in the left panel of Figure \ref{fig:mBH_fbar},
where the three solid lines indicate running medians for $0.1$-dex
$M_{200}$ bins.  We expand this analysis to the $L^*$ sample showing
dashed lines for $M_{200} <10^{12.0}$ and $>10^{12.3}\ \msolar$.  The
link between $M_{\rm BH}$ and $f_{\rm CGM}$ is muddled without
knowledge of $M_{200}$, because a small change in halo mass results in
a different $f_{\rm CGM}(M_{\rm BH})$.

The right panel of Fig. \ref{fig:mBH_fbar} considers the halo
potential, by plotting $f_{\rm CGM}$ as a function
$E_{\rm BH}/E_{\rm bind}$ in the left subpanel.  The running medians
for the same five halo mass bins show much more overlap than in the
left panel, which indicates that CGM ejection depends on the ability
of the integrated black hole energy to overcome the binding energy of
the halo.  $E_{\rm BH}/E_{\rm bind}$ is a more fundamental scaling
that controls for the halo mass dependence of gravitational potential
depth.  The colour scale of the data points indicates $M_{\rm BH}$
closely tracks this quantity with the only difference being the
$E_{\rm bind}$ in the denominator.  Four $E_{\rm BH}/E_{\rm bind}$
quartiles are indicated by black dots with x-axis error bars
indicating ranges and y-axis error bars indicating $1-\sigma$
dispersions of $f_{\rm CGM}$ for each quartile.  These quartiles
define our samples discussed starting in \S\ref{sec:cadence}.

The running medians diverge at small $E_{\rm BH}/E_{\rm bind}$, which
indicates that more massive haloes with low-mass black holes retain more
baryons.  This suggests that stellar feedback likely plays a larger
role in clearing lower mass haloes than AGN feedback.
We consider the ratio of $E_{*}/E_{\rm bind}$ (right subpanel), where
$E_*$ is the estimated integrated feedback energy from stars defined
as

\begin{equation}
E_* = \epsilon_{\rm SF} M_*,
\end{equation}

\noindent with $\epsilon_{\rm SF}\approx 1.75\times 10^{49}$ erg
$\msolar^{-1}$ represents the feedback energy return per unit mass of
star formation as calculated by \citet{cra15}.  The values for $E_*$
usually exceed $E_{\rm BH}$ and are $\ga 10\times$ higher than the
binding energy of the halo for the $M^*$ sample.  Unlike for
$E_{\rm BH}$, there is not an obvious trend, although the
$E_{*}/E_{\rm bind}$ values are limited to a smaller range by the
definition of the $M^*$ sample's stellar mass range.

Stellar feedback, which heats gas particles to $10^{7.5}$ K in EAGLE,
is less efficient at clearing out the CGM than BH feedback.  Stellar
feedback leads to gas buoyantly rising into the CGM
\citep[e.g.][]{bow17,opp18b}, but this gas often cools and recycles
back onto the galaxy.  This indicates a fundamental distinction with
BH feedback, which delivers more energy per heating event.  $E_*$ is
delivered more steadily than $E_{\rm BH}$, and a greater fraction of
this energy is lost to radiative cooling.  Stellar feedback materials
can be cycled through a sequence of superwind outflows, re-accretion
onto the galaxy, and further star formation and outflows in a sequence
often termed the ``baryon cycle.''  The $E_{\rm BH}$ feedback, which
heats gas to $10^{8.5}$ K, does not suffer as much radiative losses.
Therefore, BH feedback can disrupt or at least significantly alter
this baryon cycle by curtailing the supply of re-accreting gas
available in the CGM.
 
\section{Observational indicators of the CGM baryon fraction} \label{sec:CIV}

\begin{figure*}
\includegraphics[width=0.49\textwidth]{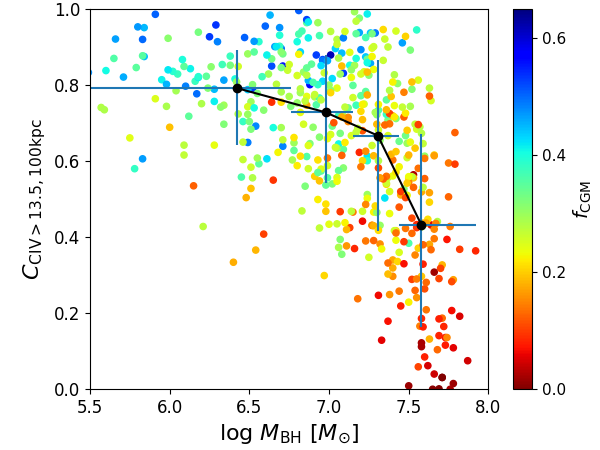}
\includegraphics[width=0.49\textwidth]{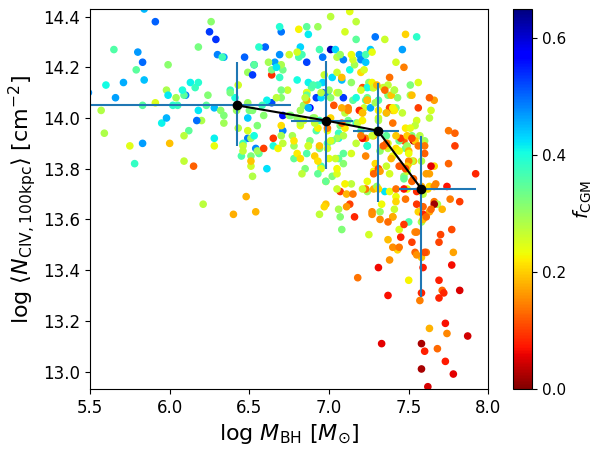}
\caption[]{{\it Left panel:} Covering fraction of $\CIV$ absorbers with column densities
  greater than $10^{13.5}\ \cms$ inside a circle of radius 100 kpc as
  function of $M_{\rm BH}$ for the $M^*$ sample at $z=0$.
  $C_{\CIV>13.5,100 {\rm kpc}}$, which is observable around local
  galaxies, is a good observational proxy for $f_{\rm CGM}$ (the
  colour of the data points), which declines the most for the highest
  mass black holes ($M_{\rm BH}>10^{7.4}\ \msolar$).  {\it Right
    panel:} The mean $\CIV$ column density, $\langle N_{\CIV, {\rm 100kpc}} \rangle$, inside the same radius 100
  kpc circle also shows a decline with $M_{\rm BH}$.  The covering fraction is a simpler statistic to obtain, because $N_{\CIV}$ measurements can be difficult or uncertain if the absorption is saturated.   }
\label{fig:mBH_fCIV100}
\end{figure*}

We have argued for a link between two physical quantities: the energy
released by the BH relative to the halo binding energy and the
fraction of a halo's baryons in the CGM.  While direct measurements of
these quantities are impossible to obtain, we focus here on the latter
quantity-- observational proxies for the CGM baryon fraction
accessible with current instrumentation.  Before continuing, we note
that we do not attempt to find observational proxies for $E_{\rm BH}$
beyond assuming a linear proportionality to $M_{\rm BH}$.  Because the
very few existing measurements of $M_{\rm BH}$ for $L^*$ galaxies are
all in the local Universe \citep{kor13}, we require that our
$f_{\rm CGM}$ proxy is also locally available.

We calculate median CGM observational measurements across four
quartiles of $M_{\rm BH}$ in the $M^*$ sample to determine how each
relates to $f_{\rm CGM}$.  The median value of the lowest (highest)
quartile of $M_{\rm BH}$ is $10^{6.42}$ ($10^{7.58}$) $\msolar$, where
$f_{\rm CGM}=0.35$ ($0.14$).  We consider several ions commonly
observed with COS, including $\HI$, $\CIV$, and $\OVI$.  Our favoured
$f_{\rm CGM}$ proxy is the $\CIV$ covering fraction of absorbers with
column densities $N_{\CIV}>10^{13.5}\ \cms$ within a circle of radius
100 kpc ($C_{\CIV>13.5, {\rm 100kpc}}$) plotted as a function of
$M_{\rm BH}$ in Figure \ref{fig:mBH_fCIV100}, left panel.  We find
that this quantity declines by nearly a factor of two from a median
with scatter of $C_{\CIV>13.5, {\rm 100kpc}} = 0.79^{+0.10}_{-0.15}$
to $0.43^{+0.24}_{-0.27}$ from the lowest to highest $M_{\rm BH}$
quartile in EAGLE.  There are three main lines of reasoning we use in
evaluating the different ions.

The first reason we favour $\CIV$ is because the 1548, 1551 \AA\
doublet is available via COS in the very local Universe ($z\la 0.01$)
and has been observed in many existing COS surveys,
\citep[e.g.][]{liang14, bordoloi14, borth15, burc15, berg18}. These
previous samples of galaxy-QSO pairs that cover $\CIV$ within the
inner CGM (R $<$ 100 kpc) of $0.1 - 1 L^*$ galaxy haloes amount to
approximately 90 sightlines in total. Future archival surveys of COS
data combined with deep galaxy spectroscopic redshifts, including the
{\it CGM$^2$} Survey (led by J. Werk), will expand the total number of
sightlines within 100 kpc covering $\CIV$ at $z < 0.1$ by at least a
factor of two.  We select $N_{\CIV}=10^{13.5}\ \cms$ as the column
density threshold because this value is easily detectable in existing
UV spectra probing the CGM. For reference, it corresponds to an
equivalent width of $\approx 100$ m\AA\ for the strong line of the
doublet at 1548 \AA, which is detectable at 3$\sigma$ in a S/N
$\approx 5-8$ COS sightline using the G160M grating.

The second reason we choose $C_{\CIV>13.5, {\rm 100kpc}}$ is that its
decline mirrors $f_{\rm CGM}$ for physically meaningful reasons.
$\CIV$ is a tracer of metals for temperatures $10^{4}-10^{5}$ K and
densities $10^{-5}-10^{-3}\ \cmc$ at $z=0$
\citep[e.g.][]{sch03,opp06,dav07,for13,rah16}.  $\CIV$ is
photo-ionized by the UV background at $T<10^5$ K for lower densities
and collisionally ionized at $\sim 10^{5}$ K for higher densities, but
these two regimes are relatively close to each other in
density-temperature phase space, which we argue make $\CIV$ a good
tracer of a well-defined region of the phase space of the diffuse CGM.
This ion still misses the majority of the CGM baryons, which typically
have $T \sim 10^4-10^7$ K and $\nh \sim 10^{-5}-10^{-2}\ \cmc$ in
MW-mass haloes.


A third reason for $C_{\CIV>13.5, {\rm 100kpc}}$ is that its decline
in covering fraction roughly follows its decline in mean CGM column
density within 100 kpc (Fig \ref{fig:mBH_fCIV100}, right panel).  The
log $\langle {N}_{\CIV, {\rm 100kpc}}\rangle /\cms$ declines from
$14.05$ to $13.72$ from the lowest to highest quartiles.  The similar
decline in $C_{\CIV}$ and $\langle {N}_{\CIV} \rangle$ (i.e. a factor
of $\approx 2$) indicates that $\CIV$ is an effective tracer of the
diffuse CGM.

The global $\CIV$ column density distributions reported in EAGLE
\citep{rah16} show reasonable agreement with observational surveys
\citep{cooksey10, burc15,danforth16}.  For the CGM, EAGLE shows very good 
agreement for $\CIV$ observed around $z\approx 2$ star-forming 
galaxies \citep{tur17}, but a comparison between $N_{\CIV}$ and 
low-$z$ EAGLE galaxies has yet to be published.  \citet{chen01} and \citet{liang14} observe strong
$\CIV$ inside 100 kpc for galaxies in the $L^*$ mass range, followed
by a rather steep decline beyond $\approx 100$ kpc. \citet{bordoloi14}
and \citet{Burchett:2016aa} studied $\CIV$ in the haloes of galaxies
over a wider mass range, finding similarly declining profiles.  We 
note here that the average $\CIV$ profiles around low-$z$ galaxies 
reported in \citet{opp18c} appear consistent with observations, and 
that the $N_{\CIV}$ measurements by \citet{Burchett:2016aa} for 
$M_*>10^{10}\ \msolar$ inside 100 kpc (their figs. 3 \& 5) are in the range of 
$\langle {N}_{\CIV, {\rm 100kpc}}\rangle$ values in Fig. \ref{fig:mBH_fCIV100}.

We list CGM measures for the lowest and highest $M_{\rm BH}$ quartiles
in Table \ref{tab:CGM_quartiles}, where we also consider $\OVI$ and
$\HI$ values repeating our three lines of reasoning used for $\CIV$.
The $\OVI$ 1032, 1038 \AA\ doublet is not locally available via COS;
however the readily detectable covering fraction,
$C_{\OVI>13.7, {\rm 100kpc}}$, shows a significant decline, which is
matched by a decline in $\langle {N}_{\OVI, {\rm 100kpc}}\rangle$
indicating it is an effective tracer of the diffuse CGM.  We discuss
the prospects and interpretation of $\OVI$ further in \S\ref{sec:OVI}.

$\HI$ is detected ubiquitously in the CGM of local galaxies via
the 1216 \AA\ Ly$\alpha$ transition
\citep[e.g.][]{tum13,borth15,kee17}.  While we find
$C_{\HI>15.0, {\rm 100kpc}}$ is a good metric for separating the
lowest and highest $M_{\rm BH}$ quartiles in Table
\ref{tab:CGM_quartiles}, $N_{\HI}=10^{15.0}\ \cms$ becomes saturated in
Ly$\alpha$.  Furthermore, $\langle{N}_{\HI, {\rm 100kpc}}\rangle$ is
3-4 orders of magnitude higher than this limit, which implies a highly
clumpy CGM tracer where most $\HI$ arises from gas with a small
filling factor \citep[][Horton et al. in prep]{for14}.

\begin{table}
\caption{EAGLE CGM values for the lowest and highest BH mass quartiles in the $M^*$ sample at $z=0$.}
\begin{tabular}{lrrr}
\hline
Metric$^1$ & Low $M_{\rm BH}$ & High $M_{\rm BH}$ & Difference \\
\hline
\multicolumn {4}{c}{}\\
log $M_{\rm BH}/\msolar$ & 6.42 & 7.58 & 1.16 dex \\
$f_{\rm CGM}$ & $0.35$      & $0.14$      & -0.39 dex \\
$C_{\CIV>13.5, {\rm 100kpc}}$ & $0.79^{+0.10}_{-0.15}$ & $0.43^{+0.24}_{-0.27}$ & -0.26 dex \\
log $\langle{N}_{\CIV, {\rm 100kpc}}\rangle/\cms$ & 14.05 & 13.72 & -0.33 dex \\
$C_{\OVI>13.7, {\rm 100kpc}}$ & $0.81^{+0.12}_{-0.17}$ & $0.32^{+0.28}_{-0.23}$ & -0.40 dex \\
log $\langle{N}_{\OVI, {\rm 100kpc}}\rangle/\cms$ & 14.03 & 13.67 & -0.36 dex \\
$C_{\HI>15.0, {\rm 100kpc}}$ & $0.64^{+0.12}_{-0.19}$ & $0.33^{+0.24}_{-0.21}$ & -0.29 dex \\
log $\langle{N}_{\HI, {\rm 100kpc}}\rangle/\cms$ & 18.76 & 17.92 & -0.84 dex \\
log $\langle{N}_{\rm H, {\rm 100kpc}}\rangle/\cms$ & 19.84 & 19.55 & -0.29 dex \\
log $\langle{N}_{\rm O, {\rm 100kpc}}\rangle/\cms$ & 16.27 & 15.87 & -0.40 dex \\
12+log (O/H)$_{\rm 100kpc}$ & 8.43 & 8.32 & -0.11 dex \\
\hline
\end{tabular}
\\
$^1$ Median values of all haloes in each quartile reported.  One-$\sigma$ scatter reported for halo covering factions.
\label{tab:CGM_quartiles}
\end{table}

Finally, we discuss the use of covering fractions calculated from a
$r=100$ kpc cylinder to determine the baryon content of a sphere
($f_{\rm CGM}$).  We calculate mean hydrogen and oxygen column
densities using the cylinder and report median CGM
$\langle {N}_{\rm H, {\rm 100kpc}} \rangle$ and
$\langle {N}_{\rm O, {\rm 100kpc}} \rangle$ where we excise the ISM
(any gas particles with non-zero SFR) for the quartiles in Table
\ref{tab:CGM_quartiles}. $\langle{N}_{\rm H, CGM}\rangle$ declines by
$-0.29$ dex while $\langle{N}_{\rm O, CGM}\rangle$ declines by $-0.40$
dex with increasing $M_{\rm BH}$, which nicely reflects the decline in
$f_{\rm CGM}$ as well as $\CIV$ and $\OVI$ covering fraction
indicators.  We can also calculate the CGM metallicity, which declines
by $-0.11$ dex from $12+{\rm log(O/H)}_{\rm CGM}=8.43$ to $8.32$.
Although the CGM metallicity declines as $M_{\rm BH}$ increases, it is
a $3.5\times$ smaller decrease than the decline of baryons within the
CGM and $2.6\times$ smaller than the baryons in a cylinder extending
to $3 \times R_{200}$ on either side of the galaxy.  We also check CGM
metallicity excluding $\nh>10^{-3.0}\ \cmc$ gas, and find the same
decline in metallicity, with $Z$ shifted by $-0.1$ dex.  The
geometrical effect of tracing baryons in a 100 kpc radius cylinder
versus a sphere of radius $R_{200}$ manifests itself in the respective
values of $\langle{N}_{\rm H, CGM}\rangle$ versus $f_{\rm CGM}$, where
the former declines by approximately one third less, because baryons
pushed out beyond $R_{200}$ remain in the cylinder along one
dimension.

\section{The evolution of the CGM and central black holes} \label{sec:cadence}

\begin{figure*}
  \includegraphics[width=0.49\textwidth]{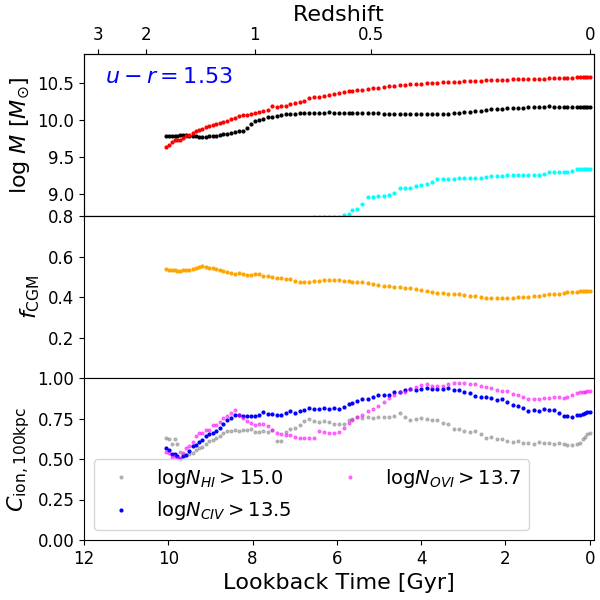} 
  \includegraphics[width=0.49\textwidth]{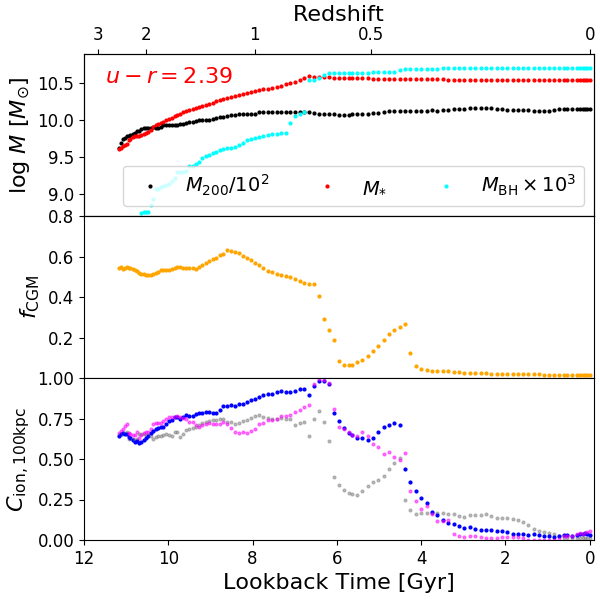}
  \caption[]{The example time histories of two secularly evolving
    galaxies with a low $M_{\rm BH}$ (left panels) and a high $M_{\rm
      BH}$ (right panels).  Upper panels show the mass evolution of
    $M_{200}$ (black), $M_{*}$ (red), and the central black hole
    (cyan).  The middle panel shows the CGM baryon content, $f_{\rm
      CGM}$, in orange.  Covering fractions of several ions, including
    $\CIV$ ($C_{\CIV>13.5, \mathrm{100kpc}}$ in
    blue) are plotted to show how these CGM observational
    proxies respond to the evolving baryon content.  The growth of the
    black hole on the right causes a reduction of $f_{\rm CGM}$ soon
    after that is reflected in $\CIV$ and other ions.  The galaxy on
    the left remains in the blue cloud at $z=0$, while the galaxy on
    the right joins the passive sequence soon after its rapid BH
    growth at $z=0.7$.  Present-day $u-r$ values are listed in the upper
    left of the upper panels.}
\label{fig:lbt_2gals}
 \end{figure*}

One of our central goals of this work is to catch the process of
baryon ejection from the CGM in the act and determine if it is a
direct result of BH feedback.  To do this, we use the rich dataset of
EAGLE snipshot outputs that follow galaxies, BHs, and the CGM in up to
155 outputs between $z=5$ and $0$ for our {\it secular} sample
comprised of 246 galaxies.  Figure \ref{fig:lbt_2gals} shows examples
of the evolution of a blue, star-forming galaxy (left) and a red,
passive (right) galaxy residing in similar mass haloes
($M_{200}\approx 10^{12.15}\ \msolar$), resulting in similar mass
galaxies ($M_*\approx 10^{10.6}\ \msolar$).  The plotting begins when
$M_*$ exceeds $10^{9.5}\ \msolar$.  We also list the final $z=0$ $u-r$
colours in the upper panels.

 The upper panels show the halo mass (black), the stellar mass
 inside an aperture of 30 kpc (red), and the central BH mass (cyan), where
 the latter are selected to be very different ($M_{\rm BH}=10^{6.3}$
 vs. $10^{7.7}\ \msolar$).  We have chosen examples in the lowest and
 highest quartiles of $E_{\rm BH}/E_{\rm bind}$ to focus on how rapid
 BH growth and the associated feedback transform the CGM, which for
 the right galaxy begins soon after $z=1$.  The middle panels show
 $f_{\rm CGM}$, which is relatively flat and slightly declining for
 the blue galaxy, but sharply dips for the eventual red galaxy first
 at $z=0.7$ as the BH grows by $2\times 10^7\ \msolar$, and then again
 at $z=0.4$ as another episode of rapid BH growth increases the mass
 by $1.3\times 10^7\ \msolar$.

 The lower panels show the time-evolving $\CIV$ covering fractions in
 blue mirroring the decline in $f_{\rm CGM}$.
 The evolution of $C_{\CIV}$ is due to several effects including 1)
 the metal enrichment of the CGM, 2) the changing ionization field
 assuming a \citet{haa01} background, and 3) the growth of the
 $R_{200}$, which is smaller than 100 physical kpc at high-$z$ and
 more than twice as large at $z=0$.  We also show $\OVI$ and $\HI$
 covering fractions within 100 kpc for these haloes in the lower
 panel, which also both respond to the rapid BH growth episodes for
 the passive galaxy.

\begin{figure}
  \includegraphics[width=0.49\textwidth]{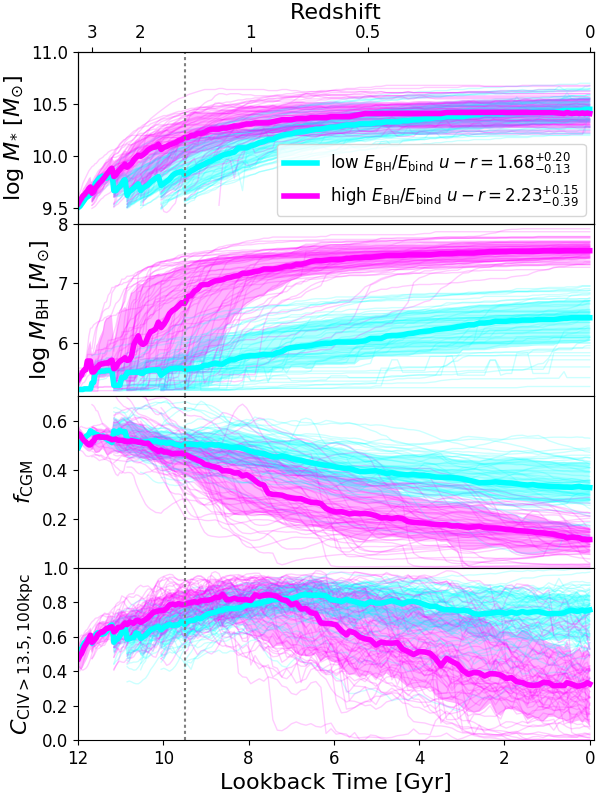}
\caption[]{The time evolution of two samples of secularly evolving galaxies, those in the
  lowest (highest) quartile of the cumulative BH feedback energy divided by gaseous halo binding energy, $E_{\rm BH}/E_{\rm bind}$, are plotted
  in cyan (magenta) as a function of lookback time.  Median histories
  are shown in solid lines with shading indicating $1-\sigma$
  spreads.  The panels from top to bottom show stellar masses, black
  hole masses (which differ by sample construction), CGM gas fractions,
  and $\CIV$ covering fractions within 100 kpc.  Gas fractions are
  similar for the two samples at $z\ga 2$, but decline more for the
  high-BH feedback sample, which leads to less star formation and
  stellar growth.  $\CIV$ traces the decline in $f_{\rm CGM}$ since
  $z=1$, although at higher redshift it shows a faster rise as enrichment
  of the CGM proceeds faster with more star formation occurring at
  these early epochs.  The grey dotted vertical line indicates the 
  redshift after which all galaxies in each sample are tracked.}
\label{fig:lbt_collate}
\end{figure}

Instead of focusing on two examples, we plot all galaxies in the
secular sample in the lowest and highest quartiles of $E_{\rm BH}/E_{\rm
  bind}$ in Figure \ref{fig:lbt_collate}. Individual galaxies are
shown with thin lines, but we show the median for each sample along
with 16-84\% shaded spreads.  All galaxies (61 in each sample) are
tracked back to $z=1.487$ (vertical dotted line), though the sample
thins out at earlier times as lower mass galaxies fall out of the
sample if $M_*<10^{9.5}\ \msolar$.  Additionally, major mergers can
occur in our sample at this epoch, which also contributes to the
sawtooth behaviour as new lower mass galaxies join the sample at
snapshot outputs corresponding to $z=$1.737, 2.012, 2.237, 2.478,
3.017, 3.528, and so on.  Even though we identify galaxies between
$z=1.487$ and $0$ without major mergers using 13 snapshots,
mergers can slip into our sample when examining high-cadence
snipshots, though at a level that does not alter the conclusions we
draw.

The stellar populations of the highest $E_{\rm BH}/E_{\rm bind}$
haloes assemble earlier than their lowest $E_{\rm BH}/E_{\rm bind}$
counterparts, resulting in a population of galaxies with much less
late-time growth and redder $z=0$ colours ($u-r = 2.23$ vs. $1.68$).
It is crucial to avoid the impression that these two sets of haloes
are the same but for their BH masses, because D19 demonstrated that
the haloes with higher mass BHs have 1) earlier formation times, 2)
higher concentrations, and 3) greater intrinsic binding energies
(calculated particle-by-particle using a matched DM-only simulation
that is devoid of baryonic effects).  Over the last 9.5 Gyr of
evolution, the high-$E_{\rm BH}/E_{\rm bind}$ haloes increase their
stellar masses by from $10^{10.2}\rightarrow 10^{10.4}\ \msolar$, while
the low-$E_{\rm BH}/E_{\rm bind}$ haloes increase from
$10^{9.8}\rightarrow 10^{10.5}\ \msolar$, a factor of $3\times$ more.
Hence, while we are focusing on secular evolution, our two sets of
haloes have fundamentally different evolutionary histories, which D19
emphasize by showing the counter-intuitive relationship that more
tightly bound haloes end up with more evacuated CGMs through greater
integrated BH feedback.

The second panel in Fig. \ref{fig:lbt_collate} following central BH
masses shows two very different tracks by design of our two samples.
The BH masses are near the seed mass at $z=2-3$ for both samples, but
the high-$E_{\rm BH}/E_{\rm bind}$ haloes have the highest rate of
growth at $z>1$.  The medians jump around at $z>1.487$ as new galaxies
are added to the sample, but some of the individual $M_{\rm BH}$ time
histories of the low-$E_{\rm BH}/E_{\rm bind}$ haloes in cyan move up
then down as more massive BHs can temporarily pass within 50 kpc of
the galaxy.  No BH masses overlap between the two samples at $z=0$ (by
sample construction) and the median BH masses are $10^{6.4}$ versus
$10^{7.6}\ \msolar$.

Next we consider the evolution of the CGM, starting with
$f_{\rm CGM}$, which begins just above $f_{\rm CGM}=0.5$ and declines
with time for both samples to their $z=0$ values ($0.33$ versus
$0.12$), which agree closely with the values reported for the two
quartiles of the $M^*$ sample (Fig. \ref{fig:mBH_fbar}, right panel).
The CGM fractions appear to respond to the growth of $M_{\rm BH}$ with
the greatest change in the high-$E_{\rm BH}/E_{\rm bind}$ haloes
occurring at epochs around $z=1$.

\begin{figure}
  \includegraphics[width=0.49\textwidth]{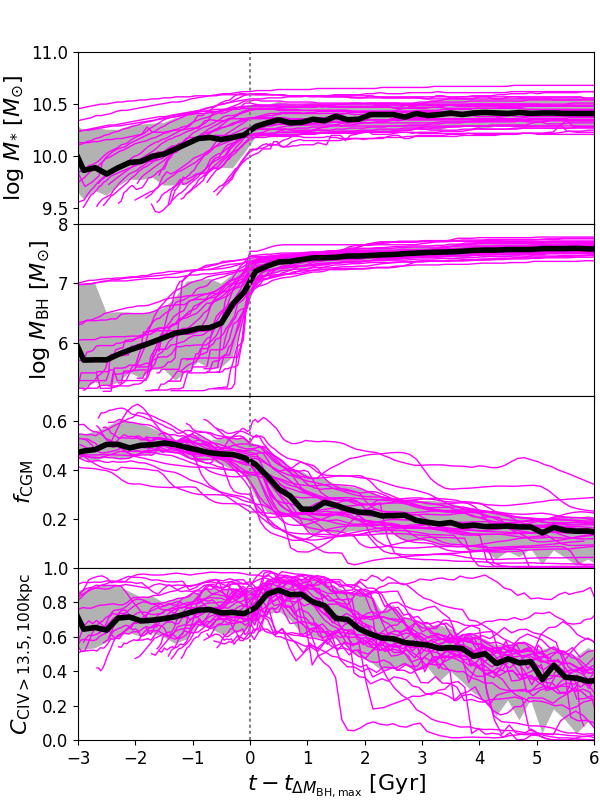}
  \caption[]{Similar to Fig. \ref{fig:lbt_collate}, but now shifted to
    the snipshot before the most rapid $M_{\rm BH}$ growth stage where the BH increases by $\Delta M_{\rm BH}>2\times 10^6\ \msolar$ in an inter-snipshot interval before
    $z=0.47$.  Grey shading indicates $1-\sigma$ ranges.  The $f_{\rm CGM}$ declines by $0.1$ within 300 Myr after
    this growth stage, which is paired with a temporary increase in $C_{\CIV}$ followed by a several Gyr decline. }
\label{fig:lbt_BHtransform}
\end{figure}

The $\CIV$ covering fraction (Fig. \ref{fig:lbt_collate}, lower panel)
also shows a response to BH growth, but it appears delayed relative to
$f_{\rm CGM}$ for the high-$E_{\rm BH}/E_{\rm bind}$
haloes as $C_{\CIV}$ declines between $z=1$
and $0.3$.  To better track how baryon clearing is linked to the BH,
we shift a set of the high-$E_{\rm BH}/E_{\rm bind}$
time histories to the snipshot interval with
the largest total BH growth, $t_{\Delta M_{\rm BH, max}}$.  Figure
\ref{fig:lbt_BHtransform} plots these shifted time histories for
galaxies with BH growth episodes at least $2\times 10^6\ \msolar$
between snipshots separated by $< 100$ Myr and
at $z>0.47$ to allow at least 5 Gyr of subsequent evolution.  The
sample contains 31 galaxies, although the trends we now discuss are
only slightly weaker if we include the other 30 galaxies.

Very little stellar assembly occurs after $t_{\Delta M_{\rm BH, max}}$
(upper panel) as the galaxy's $M_*$ essentially flatlines in response
to the biggest BH growth episode in the galaxy's history (second
panel).  The average $M_{\rm BH}$ increases by roughly $5$-fold in the
preceding 500 Myr, though our sample is not uniform further back as
$t_{\Delta M_{\rm BH, max}}$ can occur at high redshifts with little
further history.  We see a clear decline in $f_{\rm CGM}$ (third
panel) at $t_{\Delta M_{\rm BH, max}}$ indicating that there is indeed
a causal link with BH growth.  This significant growth episode of the
BH leads to a reduction of $f_{\rm CGM}$ by a factor of nearly two
(from $0.43$ to $0.23$) within 1 Gyr. This is the most clear
indication that an active, rapid BH growth episode can clear much of
the CGM.

However, as expected from Fig. \ref{fig:lbt_collate},
$C_{\CIV>13.5, 100 {\rm kpc}}$ does not respond as dramatically to the
BH growth, and in fact first jumps up from $0.76$ to $0.87$ in the 500
Myr after the BH growth episode.  Despite the initial bump, the
covering fraction has fallen to $0.49$ by $4$ Gyr.  While we argue
$\CIV$ is a good proxy for $f_{\rm CGM}$ in \S\ref{sec:CIV}, the
pathway to get there is complicated.  $C_{\CIV}$ increases in response to
BH feedback before declining, while $f_{\rm CGM}$ declines straight
away.
  
\begin{figure}
  \includegraphics[width=0.49\textwidth]{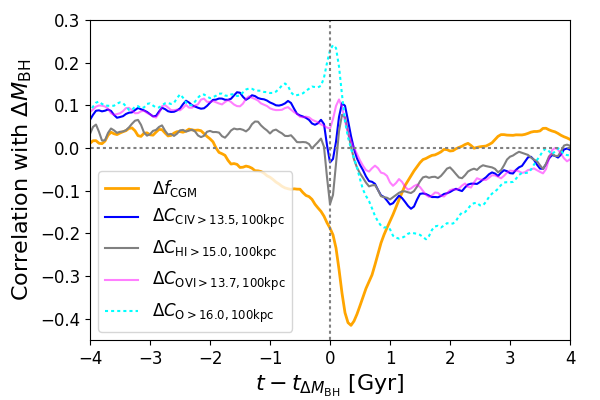}
  \caption[]{The relationships between the change in CGM quantities and BH growth as a function of time lag between BH growth at $t_{\Delta M_{\rm BH}}$ and the CGM quantity.  Average Pearson correlation coefficients, $\rho$, are plotted for all haloes in the highest $E_{\rm BH}/E_{\rm bind}$ quartile.  The negative values of $\rho$ plotted for $\Delta f_{\rm CGM}$ indicate CGM evacuation peaks 300 Myr after BH growth, and represents a direct indication that CGM gas fractions
    respond negatively to BH growth.  CGM ion covering
    fractions decline as well, but with a longer lag of 0.5-2.5 Gyr
    after a BH growth episode, and actually increase in the first 100 Myr
    after BH growth due to the ejection of ISM gas.  BH feedback initially drives metal enrichment as
    traced by the total oxygen covering fraction (dotted cyan line) before
    it declines after 400 Myr. }
  \label{fig:corr_dt_dmbh}
\end{figure}

To disentangle the connection between BH growth, $f_{\rm CGM}$, and
our covering fraction tracers, we apply a time series analysis where
we correlate all tracked BH growth on inter-snipshot intervals
($33-125$ Myr) with any change in CGM properties. For each
$E_{\rm BH}/E_{\rm bind}$ halo time history in the
high-$E_{\rm BH}/E_{\rm bind}$ secular sample, we calculate a series
of Pearson correlation coefficients, $\rho$, between
$\Delta M_{\rm BH}$ and the change in either $f_{\rm CGM}$ or a
covering fraction using all available snipshots.  If we take the time
series of $\Delta M_{\rm BH}$ and calculate its $\rho$ with the
$\Delta f_{\rm CGM}$ time series, we expect a negative value if the BH
nearly immediately ejects baryons from the CGM.  This is seen in
Figure \ref{fig:corr_dt_dmbh} when we consider the correlation with
$\Delta f_{\rm CGM}$ at $t=t_{\Delta M_{\rm BH}}$ (thick orange line
intersecting with vertical dotted grey line), which shows the mean
$\rho$ for all high-$E_{\rm BH}/E_{\rm bind}$ haloes.  This data point
alone does not necessarily indicate that the current BH growth episode
is responsible for the decline in $f_{\rm CGM}$, because BH growth
episodes are clustered in time (Fig. \ref{fig:lbt_BHtransform}) and
previous BH feedback could drive baryons out of the CGM.

We wish to identify the evolution in the correlation between CGM
quantities and BH growth, therefore we apply time lag shifts of a CGM
quantity time series relative to the BH growth time series, and plot a
series of correlation coefficients as a function of time lag in
Fig. \ref{fig:corr_dt_dmbh}.  For example, the curves crossing at
$t-t_{\Delta M_{\rm BH}}=1$ Gyr indicate the correlation of the change
in CGM quantities with the BH growth activity 1 Gyr earlier.  We plot
the mean $\rho$ values of all haloes at each time lag creating a
continuous line, demonstrating the typical time-evolving relationship
between a CGM variable and the BH growth.  The time lag evolution of
the correlation between $\Delta f_{\rm CGM}$ and $\Delta M_{\rm BH}$
value reaches its nadir 300 Myr after $t_{\Delta M_{\rm BH}}$,
indicating that CGM ejection peaks at this timescale after BH growth.
In fact, the correlation coefficient becomes sharply more negative
right after $t_{\Delta M_{\rm BH}}$ indicating an immediate
acceleration of baryon ejection, likely through shocks propagating
through the CGM at supersonic speeds.  The overall sharp dip in the
orange line is the clearest indication that EAGLE BHs are responsible
for ejecting their CGMs in short order ($\la 100$ Myr).  The negative
$\rho$ values before $t_{\Delta M_{\rm BH}}$ are due to BH growth
episodes (AGN activity) being clustered in time.  The absolute values
of $\rho$ are less important than the relative values, but given that
most $\Delta M_{\rm BH}$ values are very small, the mean Pearson
$\rho=-0.42$ at $t-t_{\Delta M_{\rm BH}}=300$ Myr is very strong.

We apply the same time series analysis to the covering fractions
focusing on $C_{\CIV>13.5, 100 {\rm kpc}}$ (solid blue line).  The
response to the the BH growth is more complicated.  $C_{\CIV}$ was
growing before $t_{\Delta M_{\rm BH}}$ indicating CGM metal
enrichment, but does respond with a sharp drop at
$t_{\Delta M_{\rm BH}}$.  Yet, $C_{\CIV}$ jumps up at 200 Myr, before
turning around and declining after 400 Myr.  The greatest decline of
$C_{\CIV}$ is between 500 Myr and 2.5 Gyr, showing a delayed response
to the CGM ejection.

We also see the same general behaviour for
$C_{\HI>15.0, 100 {\rm kpc}}$ and $C_{\OVI>13.7, 100 {\rm kpc}}$, but
at differing strengths immediately after $t_{\Delta M_{\rm BH}}$.
While it is very clear that $f_{\rm CGM}$ decreases in response to the
BH, the three different ions have responses that depend on metallicity
and ionization.  To marginalise out ionization, we plot the total
oxygen covering fraction above $N_{\rm O}>10^{16.0}\ \cms$,
$C_{{\rm O}>16.0, 100 {\rm kpc}}$ (dotted cyan), in
Fig. \ref{fig:corr_dt_dmbh}, which shows a rapid increase around
$t_{\Delta M_{\rm BH}}$ due to enrichment of the CGM by AGN feedback,
which ejects metal-rich ISM into the CGM.  The enrichment continues
but at a sharply decreasing rate until $\Delta C_{\rm O}$ becomes
negative after 300 Myr and remains negative for the next 3 Gyr.
Hence, the decline in ions in this time interval is indicative of
declining CGM metal content.  Finally, we note that the dip at
$t_{\Delta M_{\rm BH}}$ visible for the ion covering fraction is not
present for $C_{\rm O}$, which indicates that the initial dip seen for
the ions is due to AGN feedback changing the ionization state of the
gas (likely via temporary heating of lower ions to higher states).

The trends here are general for all 246 galaxies of the secular
sample, which show the same trends in average correlation
coefficients, but on average $2/3$rd as strong when adding in the
three quartiles of galaxies with lower $E_{\rm BH}/E_{\rm bind}$.

\section{Galaxy transformation as a result of CGM ejection} \label{sec:colors}

The final goal of our exploration is to understand how the galaxy as a
whole responds to CGM ejection driven by the black hole, since D19 
demonstrated that SFR is highly correlated with $f_{\rm CGM}$ at fixed halo mass.  
We report {\tt SKIRT} radiative
transfer-processed colours \citep{tra17}, because it furthers our goal
of providing the most readily available observable proxies.  
The $u-r$ colours are observed for local galaxies in
the Sloan Digital Sky Survey.  

\begin{figure}
  \includegraphics[width=0.49\textwidth]{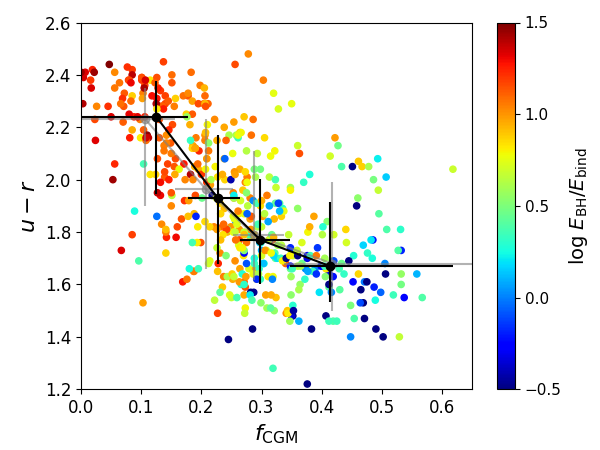}
\caption[]{Galaxy colour plotted as a function of baryonic halo gas content 
   with symbol colour indicating $E_{\rm BH}/E_{\rm bind}$ for the $M^*$ sample.  Galaxy colour clearly depends on the baryonic content of the CGM, with the
  reddest galaxies all having very evacuated haloes compared to the
  bluest galaxies.  Black points indicate running medians in quartiles
  of $u-r$ with ranges and $1-\sigma$ spreads in $f_{\rm CGM}$.  Grey points and bars show the
  same for the $L^*$ sample.}
\label{fig:color_fgas}
\end{figure}

We show our $M^*$ sample by plotting $u-r$ as a function of
$f_{\rm CGM}$ in Figure \ref{fig:color_fgas}, which shows just how
dependent galaxy colour is on CGM baryon content.  The lowest
(highest) quartile of galaxies with $f_{\rm CGM}= 0.12$ ($0.41$) have
median $u-r = 1.67$ ($2.24$).  Galaxy colours depend on their CGM
baryon content, which we have argued relies on the BH ejecting
baryons.  Hence, the content of the CGM out to at least 200 kpc is a
predictor of galaxy colour according to EAGLE.  The dark points with
error bars are shown for the $M^*$ sample
($M_{200}=10^{12.0-12.3}\ \msolar$), but we also show the trend for all
galaxies in the $L^*$ sample in grey points and error bars to show the
result is nearly same when considering all central galaxies with
$M_*=10^{10.2-10.7}\ \msolar$.

We argue that the result in Fig. \ref{fig:color_fgas} represents a new
way to think about the origin of galaxy colours.  Put another way, the
bluest (reddest) quartile of $L^*$ galaxies with $u-r = 1.59$ ($2.28$)
have $f_{\rm CGM}= 0.34$ ($0.13$) according to EAGLE.  For a more
in-depth analysis of how red sequence galaxy colours arise in EAGLE, we refer
the reader to \citet{corr19}, who used the same snipshots to
determine the timescale and cause of galaxies crossing the green
valley to the red sequence \citep[see also][]{wright19}.  They looked
at both centrals and satellites, and integrated high-cadence colour
and morphology tracking considering environmental effects, AGN
feedback, and morphological transformation as pathways for all red
sequence galaxies above $M_*=10^{10}\ \msolar$.  The most relevant
result for us is their calculation of the time between the last time a
red sequence galaxy entered the green valley and the peak black hole
growth timestep, $t_{\Delta M_{\rm BH, max}}$.  Their Figure 10 shows
a practically instantaneous process ($< 1$ Gyr centred on $0$ Gyr) of
a central galaxy entering the green valley in response to the largest
BH growth phase in a galaxy's history.  This agrees with our finding
that the gas supply from the CGM is also disrupted nearly
instantaneously through baryon clearing as shown in
Figs. \ref{fig:lbt_BHtransform} and \ref{fig:corr_dt_dmbh}.  Davies et
al. (in prep) will consider the physical mechanisms
of BH feedback impacting the CGM and transforming galaxy colours and SFRs.

\begin{figure}
  \includegraphics[width=0.49\textwidth]{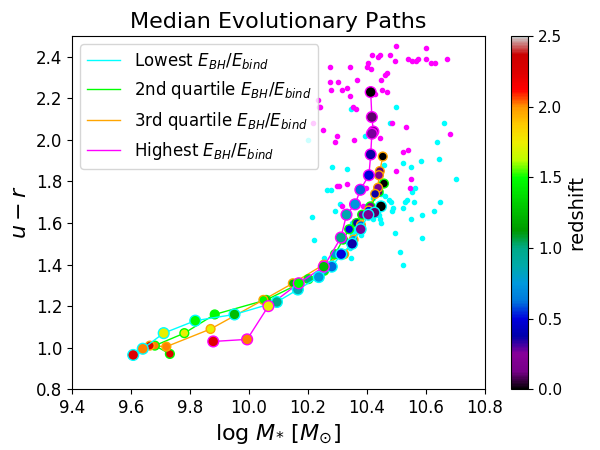}
\caption[]{Median $u-r$ colour and median stellar mass evolution of galaxies in the
  secular sample divided into $E_{\rm BH}/E_{\rm bind}$ quartiles.  The three lower quartiles have similar evolutionary paths, 
  but the highest quartile follows a divergent path that is redder since $z=1$.  
 The cyan (magenta) data points show individual $z=0$ galaxies for the  low-$E_{\rm BH}/E_{\rm bind}$  
  (high-$E_{\rm BH}/E_{\rm bind}$) quartile.}
\label{fig:color_evolution}
\end{figure}

We plot the median colour-$M^*$ evolutionary paths of the four
quartiles of the secular sample sorted by $E_{\rm BH}/E_{\rm bind}$ in
Figure \ref{fig:color_evolution}.  The paths are fundamentally
different between the lowest and highest $E_{\rm BH}/E_{\rm bind}$
quartiles, for which we plot the individual $z=0$ values in cyan and magenta,
respectively.  First, the highest $E_{\rm BH}/E_{\rm bind}$ haloes,
which predominantly end up on the red sequence, show a divergence in
colour from the other three paths going back to $z=1$.  These galaxies
are predominantly morphologically elliptical \citep{corr17}, have
their most intense black hole growth stages centred at $z\approx 1$
(Fig. \ref{fig:lbt_collate}), and form their stellar populations
earlier at the centres of haloes that collapsed earlier (D19).

Second, the other three quartiles have comparatively similar median
evolutionary paths resulting in colours that overlap more with the
blue cloud.  This is despite the fact that that each quartile of
$E_{\rm BH}/E_{\rm bind}$ has progressively lower values for $f_{\rm
  CGM}$.  This indicates that secular AGN transformation to the red
sequence requires a threshold energy of $E_{\rm BH}/E_{\rm bind}\sim
10$ for EAGLE MW-like haloes.

Finally, there exists significant scatter in $z=0$ colours across the
different quartiles, which is also apparent in Figure
\ref{fig:color_fgas}.  Even though $u-r$ colour is strongly correlated
with BH feedback history, colours alone are not a direct predictor of
the central black hole mass.  We have not considered morphology,
larger-scale environment, nor major mergers for our galaxies, which
are essential for describing all pathways of galaxy transformation.
We are also avoiding the use of dust-free, intrinsic colours, which
while not directly observable would reduce the scatter and produce
clearer trends compared to our use of colours that include the effects
of dust \citep{tra16}.

\begin{figure}
  \includegraphics[width=0.49\textwidth]{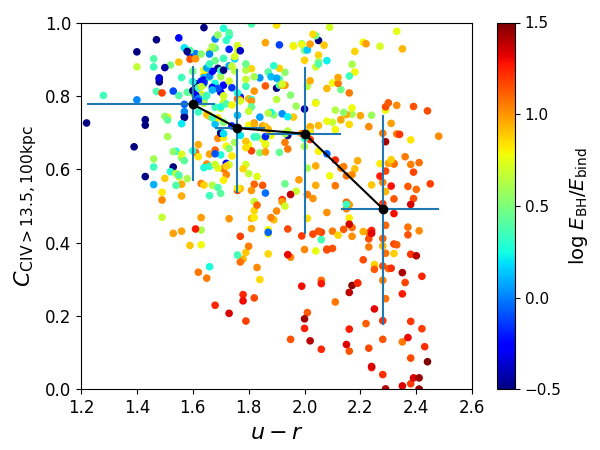} 
\caption[]{The $\CIV$ covering fraction plotted against galaxy colour
  and coloured by $E_{\rm BH}/E_{\rm bind}$ for the $M^*$ sample.
  Error bars use the same format as in Fig. \ref{fig:mBH_fCIV100}.  An
  observational version of this plot of this should be possible in the future, but the
  scatter is significantly reduced if $C_{\CIV}$ is plotted against $M_{\rm BH}$ instead (cf. Fig. \ref{fig:mBH_fCIV100}). }
\label{fig:color_fCIV}%
\end{figure}

Our approach is to provide the most direct observables to create a
plot like Figure \ref{fig:color_fCIV} using observed galaxy-absorber
pairs.  Using the same format as Fig. \ref{fig:mBH_fCIV100}, we show
points coloured by $E_{\rm BH}/E_{\rm bind}$ and medians with
$1-\sigma$ spreads on $C_{\CIV}$ and $u-r$ ranges for the $M^*$
sample.  These two galaxy-CGM observables shows a weaker correlation
than a similar plot with $M_{\rm BH}$ replacing $u-r$
(Fig. \ref{fig:mBH_fCIV100}).  The point of showing this plot is that
1) it can be created by combining existing observations \citep{burc15}
with future archival COS surveys ({\it CGM$^2$}, which would rely on
using galaxies up to $z=0.5$), and 2) we predict that obtaining
$M_{\rm BH}$ will result in a tighter correlation than colour.  We are
hopeful that the predicted link between the BH and the CGM can be
tested in the near future by collecting $\CIV$ sight lines around
nearby $L^*$ galaxies for which $M_{\rm BH}$ determinations exist.

\section{Discussion} \label{sec:discussion}

\subsection{How $\OVI$ traces the CGM} \label{sec:OVI}

In \S\ref{sec:CIV} we argued that $\CIV$ is the most promising proxy for
$f_{\rm CGM}$.  In this section we discuss $\OVI$ as a
sensitive probe of CGM baryon content even if this ion is not
currently locally available via COS.  Considerable debate has emerged on the
nature and origin of $\OVI$ in the COS-Halos Survey after it was shown
to be strong around blue, star-forming and weaker around red, passive
$z\approx 0.2$ galaxies \citep{tum11}.  \citet{opp16} used EAGLE zooms to
argue that $\OVI$ declined around passive galaxies, not due to baryon
ejection, but instead owing to the ionization effect of these galaxies
living in much more massive haloes ($M_{200}\sim 10^{13}\ \msolar$)
with virial temperatures $>10^6$ K.  \citet{nel18b} used the Illustris-TNG simulations
with an updated BH feedback scheme \citep{weinberger17} to show that
$\OVI$ is reduced significantly by BH feedback in the $M^*$ halo mass
regime.  Illustris-TNG also shows a stark decline in $f_{\rm CGM}$,
falling from $\approx 0.6$ to $\approx 0.2$ for centrals with
$M_*=10^{10.3}$ to $10^{10.7}\ \msolar$ \citep[][their
fig. 20(ii)]{nel18b}, which indicates baryon ejection by the BH.

Given that we are using the main EAGLE simulation, which shows the
same trends as the zooms \citep{opp16}, we know that $\OVI$ declines
for multiple reasons: BH feedback evacuating the halo (D19; this
paper) at $M_{200}\sim 10^{12}\msolar$ and the virial temperature effect at higher halo masses \citep{opp16}.  For EAGLE,
median $f_{\rm CGM}$ increases monotonically with halo mass, while
median $f_{\rm CGM}$ declines as a function of increasing halo mass
over the $M^*$ mass range in Illustris-TNG \citep[][their fig. 4,
measured at a smaller radius]{pil18}.  This indicates that BH ejection
is not as aggressive in EAGLE, but it is still enough of a factor to
reduce $f_{\rm CGM}$ significantly at fixed halo mass.  A COS
observational survey targeting $\CIV$ or $\OVI$ around star-forming
and passive galaxies at fixed halo mass could better isolate the effect
of BH evacuation of the CGM.  It remains unclear if COS-Halos passive
galaxies live in similar mass haloes as their star-forming
counterparts, but the passive galaxy stellar masses are mostly
higher than $M_*=10^{10.7}\ \msolar$ \citep{wer12}.

\subsection{Consequences of black hole feedback efficiency} \label{sec:BHeff}

The energy efficiency of BH feedback has important consequences for a
galaxy and its CGM.  In EAGLE, the efficiency of the conversion of rest-mass energy of
material accreting onto the BH to feedback energy is
$\epsilon_{\rm BH}\equiv \epsilon_r \epsilon_f=1.5\%$ \citep{boo09}.
The Illustris-TNG simulations use a two-mode BH feedback scheme where
their thermal mode has $\epsilon_{\rm BH}=2\%$ and is used for high
accretion rates, but their kinetic mode has $\epsilon_{\rm BH}=20\%$
and is used for low accretion rates \citep{weinberger17}.  The
kinetic mode operates primarily at later times and for later growth of
the BH, and represents high efficiency jet feedback, which appears to be responsible for reducing $\OVI$ and $f_{\rm CGM}$
for passive $M^*$ haloes in Illustris-TNG \citep{nel18b}.  The ROMULUS25 simulation
uses a thermal BH feedback scheme with much lower efficiency,
$\epsilon_{\rm BH}=0.2\%$ \citep{tremmel17}.  \citet{sanchez19}
demonstrated that the addition of BH feedback increases CGM metal
enrichment, likely because it is not powerful enough to clear the CGM
of baryons (their fig. 9), and may have more dynamical similarities to
EAGLE's stellar feedback scheme.

Galaxy colours for both the EAGLE and Illustris-TNG 100 Mpc volumes
show good agreement with observational data, but there are
inherent differences.  The EAGLE {\tt SKIRT} dust-processed colours,
which we use in our analysis, display a colour bimodality for $L^*$
galaxies with slightly more blue peak galaxies than observed
\citep[][their fig. 7]{tra17}.  The Illustris-TNG colours, applying
‘resolved’ dust attenuation based on the neutral gas and metal
distributions within galaxies, display a colour bimodality with
slightly more red peak galaxies for massive $L^*$ galaxies than
observed \citep[][their fig. 1]{nel18a}.  The red colours of
Illustris-TNG $L^*$ galaxies are likely more attributable to efficient BH
feedback, while the red colours for EAGLE $L^*$ galaxies are likely more
influenced by dust effects.  The ROMULUS25 volume is far smaller, but
it appears that their passive galaxy fraction is lower than for EAGLE and
Illustris-TNG \citep[][fig. 1]{sanchez19}.  Taken together, the
feedback efficiency of the BH central engine appears to be linked to
both the CGM gas content and the galaxy's stellar assembly and colour across multiple simulations.

\subsection{The Milky Way CGM-black hole connection} \label{sec:MilkyWay}

The most recent estimates of the MW halo mass are closer to
$M_{200}=10^{12.0}\ \msolar$ \citep[e.g.][]{eadie19}, though we
consider the range of possibilities to be
$10^{12.0}-10^{12.3}\ \msolar$ \citep[e.g.][]{battaglia05}.  Sgr A* has
a mass of $4\times 10^6\ \msolar$ \citep{boehle16}, which puts our
galaxy below $M_{\rm BH}/M_{200}<10^{-5}$ and suggests it is most
similar to galaxies in the lowest $E_{\rm BH}/E_{\rm bind}$ quartile.
We use the \citet{bre18} estimates for the total CGM mass, and assume
their values of $M_{\rm CGM,T<10^5 K}=10^{10}\msolar$ and
$M_{\rm CGM, T\geq 10^5 K}=6\times 10^{10}\ \msolar$, and plot the MW's
$f_{\rm CGM}$ for four assumed halo masses in Figure
\ref{fig:m200_fgas}.  The MW's gaseous halo appears to agree with the EAGLE
model, especially if its $M_{200}$ is indeed closer to $10^{12.0}$ than
$10^{12.3}\ \msolar$.  The MW contains an under-massive BH and a higher
$f_{\rm CGM}$ compared to average MW-mass haloes.  Further constraints
on the MW CGM mass, especially its hot component at larger radii, will
allow more accurate explorations of our halo's BH-CGM link.  While
halo gas fractions of clusters appear over-estimated by the EAGLE
model \citep[][C-EAGLE]{bar17}, the $f_{\rm CGM}$ gas fractions of
MW-mass haloes are observationally unconstrained.

Our results suggest that the MW may be on the precipice of an epoch of
rapid BH growth and CGM evacuation.  Using the sequence that
\citet{bow17} derived from EAGLE, the growth of the hot halo leads to
a reduction in the efficiency of feedback from star formation.
Ineffective stellar feedback results in gas collecting near the galaxy
centre, leading to rapid BH growth and AGN feedback
\citep[see also][]{dubois15,ang17b}, which we argue can eject a large fraction of
the baryons from our halo.  However, it must be realised that similar
mass haloes with massive BHs underwent the stage of rapid BH growth at
much higher redshift, and that the MW halo is not interchangeable with
a halo of the same mass hosting a passive galaxy.  D19 showed that
high-$M_{\rm BH}$ haloes had earlier formation times than
low-$M_{\rm BH}$ haloes.  This is also interesting for the MW, because
metal abundance patterns of the MW disc \citep{mackereth18} and the MW
globular cluster age-metallicity distribution \citep{kruijssen18}
suggest an earlier formation history than typical disc galaxies
occupying $\sim 10^{12}\ \msolar$ haloes.  Combined with the later formation 
times implied by its under-massive BH, the MW system could be atypical.  


\section{Summary} \label{sec:summary}

We analyse the largest EAGLE simulation to understand how the baryonic
content of Milky Way (MW)-mass galaxy haloes reacts to the growth of their
central supermassive black holes and how the resulting evolution
transforms the galaxies.  Our investigation leverages the high-cadence
tracking of a large sample of simulated galaxies to catch the act of
``baryon lifting'' where the BH feedback energy can eject most of the
circumgalactic baryons from the halo and transform the galaxy across
the green valley to the red sequence.  We argue that the key physical
quantity is the time integral of the total energy released by the BH
divided by the binding energy of the gaseous halo
($E_{\rm BH}/E_{\rm bind}$).  We attempt to relate these physical
characteristics to accessible observational proxies of 1) the
cumulative BH feedback energy, through $M_{\rm BH}$, 2) the gaseous
content of the CGM, through ion column densities and covering fractions,
and 3) the star formation history of the galaxy, through $u-r$
colours.

This work is an extension of \citet{davi19a} who identified
$M_{\rm BH}$ as being highly anti-correlated with the halo gas
fractions, $f_{\rm CGM}$, in EAGLE.  Here, we mainly focus on
present-day $L^*$-mass galaxies ($M_*=10^{10.2-10.7}\ \msolar$)
residing in MW-mass haloes ($M_{200}=10^{12.0-12.3}\ \msolar$) that
show a great diversity not only in galaxy colours, but also in
$M_{\rm BH}$ and $f_{\rm CGM}$.  The main results of our investigation
connecting the scales of BHs, galaxies, and the CGM are as follows:

\begin{itemize}

\item{While $f_{\rm CGM}$ is strongly anti-correlated with
    $M_{\rm BH}$ at a given halo mass, the anti-correlation with 
    $E_{\rm BH}/E_{\rm bind}$ appears to be more fundamental for
    $M_*=10^{10.2-10.7}\ \msolar$ galaxies.  The median value of
    $f_{\rm CGM}$ is $0.35$ ($0.14$) in the lowest (highest) quartile
    of $E_{\rm BH}/E_{\rm bind}$, which has a median value of $E_{\rm BH}/E_{\rm bind}=1.2$
    ($15$). (\S\ref{sec:CGM_BH}, Fig. \ref{fig:mBH_fbar})}
\item{We explore covering fractions for several ions observable (by COS),
    including $\HI$, $\CIV$, and $\OVI$, and argue that the
    $N_{\CIV}>10^{13.5}\ \cms$ covering fraction within 100 kpc of the
    galaxy, $C_{\CIV}$, is most easily obtainable for local galaxies
    where BH mass estimates are most readily available.  $C_{\CIV}$ mirrors
    the trend of $f_{\rm CGM}$ and declines from $0.79$ to $0.43$ from
    the lowest to the highest $E_{\rm BH}/E_{\rm bind}$ quartile.  $\OVI$ is
    also an effective $f_{\rm CGM}$ proxy, but is not locally
    available with COS. (\S\ref{sec:CIV}, Fig. \ref{fig:mBH_fCIV100})
    }
  \item{High-cadence tracking of a subset of our galaxies indicates a
      causal link between $M_{\rm BH}$ and $f_{\rm CGM}$.
      $f_{\rm CGM}$ responds to BH growth on a cosmologically very
      short, $<100$ Myr, timescale.  Ion covering fractions take
      longer to decline in response to episodes of BH growth
      ($0.5-2.5$ Gyr), which is in part due to AGN-driven metal
      transport from the ISM to the
      CGM. (\S\ref{sec:cadence}, Figs. \ref{fig:lbt_collate}, \ref{fig:lbt_BHtransform}, \ref{fig:corr_dt_dmbh})}
  \item{The $u-r$ colours, calculated using the {\tt SKIRT} radiative
      transfer dust-reddening model, have values of
      $2.23^{+0.15}_{-0.38}$ and $1.68^{+0.20}_{-0.14}$ for the
      highest and lowest $E_{\rm BH}/E_{\rm bind}$ quartiles.  Hence,
      haloes having undergone significant secular BH growth are more
      likely to be red sequence galaxies while their low-$M_{\rm BH}$
      counterparts likely remain in the blue cloud.  However, the
      significant dispersion in $u-r$ values indicates that
      $E_{\rm BH}/E_{\rm bind}$ alone is not a good predictor of
      colour. (\S\ref{sec:colors}, Figs. \ref{fig:color_fgas}, \ref{fig:color_evolution}, \ref{fig:color_fCIV})}

  \item{The MW itself has a low $E_{\rm BH}/E_{\rm bind}$
      ratio, calculated using the mass of Sgr A*, which would indicate
      that it should have a high $f_{\rm CGM}$ for its halo mass.
      Estimates for the observed gas mass of the MW, while
      highly uncertain \citep[e.g.][]{bre18}, suggest that it retains
      more gas than the average halo if its $M_{200}$ is
      $\approx 10^{12.0}\ \msolar$. (Fig. \ref{fig:m200_fgas})}
  \item{Although the time-integrated stellar feedback energy released
      is greater than the BH feedback released in most EAGLE MW-like
      haloes, the more gradual release of stellar energy to the CGM
      does not clear the halo, but maintains a cycle of accretion,
      feedback, and re-accretion as the galaxy evolves along the
      star-forming sequence.  The energy release due to rapid growth
      of the BH can disrupt and fundamentally change the cycle of
      baryons between the CGM and galaxies, leaving lasting impacts on
      both.}
    
\end{itemize} 

We put forth that a fundamental pathway for secular galaxy transformation
involves a three-step sequence: 1) the formation of a hot halo, 2) the
rapid growth of the BH, and 3) the lifting by AGN feedback of the
baryonic halo curtailing the supply of fuel for star formation.  The
first two of these processes were linked in EAGLE by \citet{bow17},
who argued that the hot halo prevents effective SF feedback from
buoyantly rising into the CGM, leading to increased accretion onto and
rapid growth of the central BH.  \citet{davi19a} revealed the
inverse correlation between $M_{\rm BH}$ and $f_{\rm CGM}$ in EAGLE, suggesting
a causal link between the BH and the removal of a significant portion
of the gas from the halo, which reduces CGM accretion and galactic
star formation.  We uncover this causal link in a set of galaxies that
rapidly grow their BHs, eject baryons from their haloes, and transform their
colours.

Our analysis contends that the efficiency with which a BH couples its
feedback energy to the CGM is essential for understanding the process
of secular galaxy transformation.  The relationship between a galaxy's stellar assembly, its halo gas supply, and its central BH should be a focus of CGM-oriented studies going forward.  We emphasise the need for observational campaigns targeting local galaxies, where non-AGN BH masses and dynamical halo masses are or will become available.

 
\section*{acknowledgements}

The authors would like to thank Akos Bogdan, Lars Hernquist, Ian McCarthy, Dylan Nelson, Erica Nelson, Nicole Sanchez, John Stocke, and Rainer Weinberger for stimulating conversations and exchanges that contributed to this manuscript.  Support for BDO was provided through the NASA ATP grant NNX16AB31G and NASA {\it Hubble} grant HST-AR-14308.  JJD acknowledges an STFC doctoral studentship. RAC is a Royal Society University Research Fellow. JKW acknowledges support from a 2018 Sloan Foundation Fellowship.  The  study  made  use  of  high performance computing facilities at Liverpool John Moores University, partly funded by the Royal Society and LJMU’s Faculty of Engineering and Technology, and the DiRAC Data Centric system  at  Durham  University,  managed  by  the  Institute  for Computational Cosmology on behalf of the STFC DiRAC HPC Facility  (http://www.dirac.ac.uk).  This equipment was funded by BEIS capital funding via STFC capital grants ST/K00042X/1, ST/P002293/1, ST/R002371/1 and ST/S002502/1, Durham University and STFC operations grant ST/R000832/1. DiRAC is part of the National e-Infrastructure.

\bibliographystyle{mnras}
\bibliography{BH_CGM.v5.bib}

\end{document}